# Model-in-the-Loop (MILO): Accelerating Multimodal AI Data Annotation with LLMs

Yifan Wang[1], David Stevens[1*], Pranay Shah[1*], Wenwen Jiang[2], Miao Liu[2], Xu Chen[3], Robert Kuo[1], Na Li[1], Boying Gong[3], Daniel Lee[1], Jiabo Hu[1], Ning Zhang[2], Bob Kamma[1]

*equal contribution, correspondence: {wangyifan, vkamma}@meta.com

[1]Labeling Platform, Meta

[2]GenAI, Meta

[3]Central Applied Science, Meta

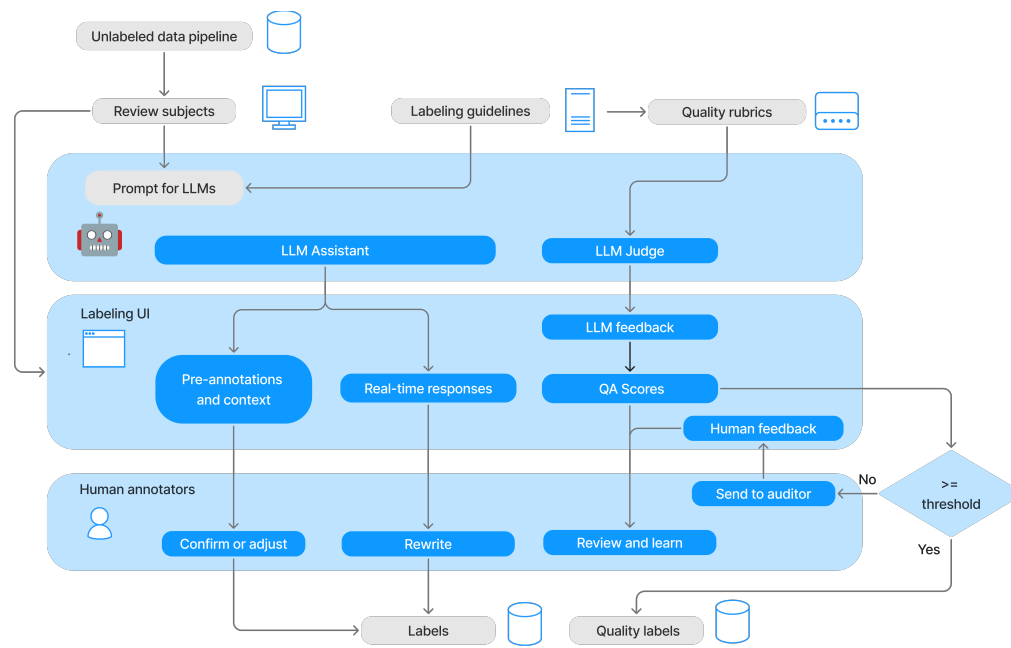

Figure 1: Model-the-in-loop (MILO) framework for data annotation systems

The growing demand for AI training data has transformed data annotation into a global industry, but traditional approaches relying on human annotators are often time-consuming, labor-intensive, and prone to inconsistent quality. We propose the Model-in-the-Loop (MILO) framework, which integrates AI/ML models into the annotation process. Our research introduces a collaborative paradigm that leverages the strengths of both professional human annotators and large language models (LLMs). By employing LLMs as pre-annotation and real-time assistants, and judges on annotator responses, MILO enables effective interaction patterns between human annotators and LLMs. Three empirical studies on multimodal data annotation demonstrate MILO's efficacy in reducing handling time, improving data quality, and enhancing annotator experiences. We also introduce quality rubrics for flexible evaluation and fine-grained feedback on open-ended annotations. The MILO framework has implications for accelerating AI/ML development, reducing reliance on human annotation alone, and promoting better alignment between human and machine values.

## 1 INTRODUCTION

AI data annotation refers to the process of labeling or tagging data with relevant information, such as text, tags, or classifications. This process enriches the data with context and meaning, thereby facilitating the training, evaluation, and adversarial testing of AI or machine learning (ML) models. Typically, the annotation tasks are performed manually by human annotators, whose expertise is important for ensuring the accuracy and reliability of the data. Today, data annotation has evolved into a global industry[37,45], leading to the emergence of human annotators as both gig workers and professionals. Data annotation for large-scale industry models remains a time-consuming, labor-intensive process, where achieving consistent quality can be challenging. With the advancement in foundation models including Large Language Models (LLMs), recent research has explored the potential of models to either replace or augment human annotators, aiming to improve efficiency and reduce the dependency on human labor. Despite these advancements, the complexity, subjectivity, and diversity of data often require domain expertise, meaning that resource-intensive manual work is still indispensable.[3] Consequently, it is important to design systems that leverage both human annotators and models throughout the annotation process, namely through human-model collaboration [18,23].

Previous research has focused on comparing the performance of LLMs with that of crowdsourced workers on text classification[1,13,16], with limited attention given to how these models could improve the productivity of professional annotators in large-scale production tasks. Furthermore, these studies typically applied models in limited roles, such as confidence filters or additional labelers in multi-review settings, without fully leveraging the possibilities of enhancing human annotation through strategic human-computer interaction (HCI) design principles.

To address these limitations, this paper introduces the Model-in-the-Loop (MILO) framework, designed to integrate AI/ML models into the human annotation process and improve annotation efficiency and quality. The MILO framework is model agnostic, employing existing models, including LLMs, which are under development or previously trained for different tasks. Our research objectives are threefold:

- To develop an interactive framework for human-model collaboration throughout the data annotation lifecycle, establishing design principles and identifying specific model applications.
- To introduce a methodology for evaluating subjective quality criteria that capture human preferences and values in data annotation, enabling more nuanced assessments of annotation quality.
- To empirically evaluate the effectiveness of the MILO framework through three studies conducted in real-world production settings with professional annotators and researchers.

The MILO framework aims to enhance annotation efficiency and quality, boost downstream model performance, and ensure better alignment between human and machine values, ultimately contributing to more efficient AI/ML model development.

## 2 RELATED WORK

### 2.1 Expanding datasets with models

In machine learning, model assistance is an established practice for expanding datasets and provides cost-effective alternatives to labor-intensive annotation for fully supervised learning. For instance, semi-supervised learning expands the dataset by using the model's predictions to label the unlabeled data.[5,7] Meanwhile, synthetic data generation creates artificial datasets from scratch, which is useful when real data is unavailable, lacks diversity, or privacy is a concern.[24,52] Weak supervision, using noisy data to train models, is useful when fully labeled datasets are impractical, allowing initial training on less accurate data and refinement over time.[27,32,53,54]

### 2.2 Model outperforms crowdsourced annotators

Previous research has primarily focused on comparing the performance of models to that of human annotators, with the potential to replace crowdsourced human annotators. LLMs have been particularly noted for their cost-effectiveness. For instance, ChatGPT exhibits superior capabilities to MTurk workers across diverse annotation tasks, encompassing relevance, stance, topics, frames detection [13], social media posts [41], and foreign language genre classification[20]. The per-annotation cost of ChatGPT is notably lower than that of crowd-workers, estimated to be approximately twenty times cheaper than MTurk as reported by Gilardi et al [13]. Similar studies also suggest that open-source large language models, including Llama[43], HuggingChat, and FLAN[48], often exceed the capabilities of crowdsourced human annotators.[1] Their growing popularity stems from their transparency, reproducibility, and enhanced data protection, allowing organizations to maintain control over sensitive data without relying on third-party APIs.

### 2.3 Model augments human annotators

Models can augment human annotators by acting as independent annotators, generating annotations for direct use in downstream model training. Specially, models can function as filters to select the most informative data for human review or verification.[11,14,47] This approach, also known as the co-annotation strategy, involves LLMs identifying subsets of data where they exhibit uncertainty. Directing these uncertain cases to human annotators can maximize the value derived from human input, aligning with active learning principles [23]. Furthermore, integrating model-generated labels with human labels has proven beneficial, particularly when models produce additional labels during the multi-review process. For instance, combining GPT-4 labels with MTurk human annotations through advanced label aggregation techniques has improved the accuracy of text annotation tasks, such as identifying sections in research paper abstracts. [16] There is a need for systems that can effectively manage and distinguish between labels generated by humans or different model versions. MEGAnno+ [18] is one such system, though it currently functions solely at the Jupyter Notebook level and does not provide a complete end-to-end solution.

### 2.4 Model assists human annotators

Models can serve as in-UI assistants for various interactive labeling tasks by providing suggestions and context. In classification tasks with large label sets, research has demonstrated that models can leverage predicted confidence scores to preselect or suggest possible labels, helping annotators narrow down the decision space and reduce handling time. [9,10] Some studies have also explored the potential of LLMs to provide additional contextual information, such as natural language explanations, which can aid human annotators in more nuanced tasks like identifying hate speech [17] or addressing visual commonsense challenges [35]. Model assistance is particularly important for computer vision tasks such as video object tracking and segmentation, which require meticulous frame-by-frame and object-by-object annotation. The Segment Anything model exemplifies the benefits of model assistance, utilizing previously trained models to auto-populate frames with masks based on initial inputs.[19,34] Without this assistance, such tasks would be prohibitively tedious and time-consuming for human annotators.

## 2.5 Model as a judge for human annotation

Certain models are specifically trained to either assess and critique responses generated by humans or other models, or rank multiple responses based on predefined criteria or error codes. To ensure their accuracy and relevance, judge models are evaluated using human preference data and automatic benchmarks, allowing them to serve as proxies for human judgment.[46,50] Such an approach is often referred to as "model-as-a-judge". Research has shown that state-of-the-art (SOTA) models like GPT-4[30] and Llama-3 70B closely align with human evaluations, demonstrating robustness and cost-effectiveness.[42] In addition, human-model collaboration has been shown to boost the quality of human-written feedback. For instance, critiques from a fine-tuned model can improve human annotators' critiques in text summarization tasks [36], while models like CriticGPT,[26] trained on human feedback, can effectively identify bugs in code review, which leads to more comprehensive human critiques when human annotators are presented with model-generated critiques.

## 2.6 Data annotation for LLM development

Modern foundation models are developed in two main stages: (1) pre-training, where the model learns from a vast amount of data, and (2) post-training, which fine-tunes the model to better follow instructions and improve specialized skills like reasoning, coding, or multilingual capabilities. [40,43,44] The post-training phase is heavily reliant on human-annotated data, through Supervised Fine-Tuning (SFT) and Direct Preference Optimization (DPO) [33]. In SFT, human annotators develop high-quality prompts that are paired with responses either generated by the model or humans. Meanwhile, DPO focuses on refining the model based on preference data, where annotators evaluate and rank various model outputs (e.g., preferring "A" over "B"). Aside from DPO, there are more complex alignment strategies such as Reinforcement Learning from Human Feedback (RLHF) [3,6,28,31,38,51], though the annotation workflow remains unchanged compared to DPO. Techniques such as Reinforcement Learning from AI Feedback (RLAIF)[21] and Constitutional AI [4] employ model-generated preferences as reward signals. These methods facilitate precise control over AI behavior with reduced dependence on human-generated labels. Human oversight is effectively maintained through a carefully curated set of examples, rules, or principles.

## 3 METHODOLOGY

Despite the growing importance of model integration in data annotation, research on model-assisted methodologies remains in its early stages. Caution is advised when relying on model signals, as their decisions can introduce biases and potentially mislead annotators. There is a risk that annotators may become over-reliant on model-generated answers, which could result in the downstream model missing out on valuable human feedback. In practice, teams often prefer human annotations over synthetic data due to the nuanced and subjective nature of annotation tasks. Therefore, while models cannot replace human annotators, the focus should shift towards fostering effective human-model collaboration. Currently, comprehensive studies that explore the design of such collaborative paradigms are lacking.

As outlined in the related work section, models play three key roles in the labeling lifecycle: serving as an assistant, labeler, or judge, depending on the interaction between human annotators and the tasks performed by the model. With the increasing prominence of LLMs, it is important to focus on annotation workflows like SFT, where human annotation is essential for optimizing model performance. Our MILO framework is bidirectional; it not only leverages LLMs to improve annotation processes but also utilizes these annotations to enhance LLM development, thereby creating a more efficient lifecycle. To this end, we propose a collaborative annotation framework involving both LLMs and human annotators, aimed at enhancing annotation efficiency and quality (Figure 1). We evaluated the efficacy of the MILO framework through a series of experiments (Section 4), conducted on our internal labeling platform.

## 3.1 Data annotation system overview

The traditional data annotation systems are comprised of several key components: human actors, input and output data, labeling user interfaces (UI), and guidelines.

**Human actors** include 1) annotators: these individuals are responsible for handling the majority of production annotation tasks; 2) auditors: as more experienced annotators, auditors are specially trained in quality assurance (QA). Due to their advanced skills and knowledge, they generally compensated at higher rates; 3) ML researchers or product owners who are responsible for drafting the annotation guidelines and quality rubrics. For simplicity, we will refer to them as researchers in this work. Typically, researchers are directly associated with the organization that owns the model development, in contrast to annotators and auditors, who are often employed by third-party vendors.

**Input data**: unlabeled data, typically in a tabular format, enters the system via an upload pipeline. The raw data is then rendered in the labeling UI as review subjects, containing various formats such as images, videos, texts, or chat conversations based on the specific annotation tasks. It is essential for all contextual information to be included to enable annotators to carry out their tasks.

**Output data**: the primary outputs of annotation tasks are labels or annotations. The content of these outputs varies depending on the task, including coordinates for keypoint and bounding box annotations, encoded masks for segmentation, text responses for text generation tasks, or categories/tags for classification tasks. Once the annotations are completed, they can be retrieved and downloaded in bulk through dedicated pipelines.

**Labeling guidelines** serve as the instruction manual for annotators, drafted by researchers or product owners. These guidelines encompass detailed instructions on how to use the labeling tools, descriptions of the annotation tasks or workflows, including the questions to be answered, required responses with their definitions, and illustrative examples. Additionally, they include quality rubrics and visual aids such as screenshots or layouts of the labeling UI.

**Labeling UI** should be designed to cater to different users. For regular annotators, the UI includes a list of review subjects and a form containing the question/options, and potentially input boxes for drafting and submitting answers. For auditors and researchers, the UI includes review subjects along with previously generated annotations, allowing them to accept or reject annotations and provide additional feedback based on quality rubrics. Once QA feedback is submitted, annotators have access to a view where they can review all the feedback received, enabling them to improve their annotation skills.

## 3.2 Quality rubrics and feedback

In tasks including text response generation and LLM chat annotation, such as those used in SFT, quality assessment is subjective, thus challenging to quantify with numerical metrics. To standardize evaluations, we introduce structured rubrics tailored to each annotation project. These rubrics establish clear standards for high-quality annotation. QA feedback, provided per annotation, identifies specific errors or assigns a categorical grade for each quality criterion. This structured feedback is essential for helping annotators understand their performance and identify improvement areas. The QA score, derived from this feedback, offers a numerical representation of annotation quality, facilitating quality monitoring and annotator performance tracking. Based on our experiences, we have identified two primary categories of rubrics:

**Point deduction rubrics** outline specific error categories and their corresponding penalty deductions. The rubric consists of $n$ possible error categories. For each annotation, there can be multiple occurrences of the same type of error. Let's denote the QA score as S, the maximum possible score as M, the penalty for each error category as $e_i$, and the occurrence of each error category as $m_i$. The QA score can be calculated using the following equation:

$$S = M - \sum_{i=1}^{n}(e_i m_i) \qquad (1)$$

**Grading scale rubrics** evaluate the quality of annotations against a set of predefined criteria, assigning grades that reflect the level of adherence to these standards (e.g. on a scale of 1-5). Each criterion is weighted to underscore its importance relative to the overall assessment. Let's denote the QA score as S, the number of criteria as $n$, the weight of each criterion as $w_i$, the numerical grade assigned to each criterion as $g_j$, and the number of grades for each criterion as $g_i$. The QA score is then computed as:

$$S = \sum_{i=1}^{n}(w_i \frac{g_j}{g_i}) \qquad (2)$$

For example, consider the criteria "helpfulness," "truthfulness," and "harmlessness" from the InstructGPT guideline [31], with respective weights of 30%, 30%, and 40%. If an auditor assigns grades of 3 for helpfulness, 2 for truthfulness, and 5 for harmlessness, the final score is $S = (0.30 \times \frac{3}{5}) + (0.30 \times \frac{2}{5}) + (0.40 \times \frac{5}{5})$, Thus, the final score $S$ is 70%.

There are two possible options for QA status, "PASSED" and "REDO." If the QA score falls below a specified threshold set by the researchers, the response status will be marked as "REDO". In such instances, both the feedback and the original annotation should be returned to the original annotator for review and learning purposes. Additionally, responses marked as "REDO" shall not be included in the download pipeline as quality labels.

### 3.3 Model-in-the-loop (MILO)

Models, like human annotators, receive identical review subjects as input. Additionally, LLMs incorporate instructions from the labeling guidelines directly into their prompts. It is important that these labeling guidelines and quality rubrics are designed in a model-driven, structured manner to minimize ambiguity and subjectiveness, ideally incorporating examples where possible.

*3.3.1 Pre-annotation LLM assistant*

Pre-annotation refers to the process of using models to automatically generate initial annotations or guidance in the labeling UI, which are then reviewed and validated by human annotators. The primary goal is to augment human annotation and support decision-making. To minimize idle time, pre-annotations are generated prior to human annotators start working. The guidance provided can include confidence scores, contextual insights, and detailed explanations for each option in classification tasks. By integrating relevant details from labeling guidelines, the assistant offers targeted suggestions and clarifications that help annotators navigate complex questions and nuances. For tasks involving extensive classification options, the assistant can also pre-select and rank top recommendations, reducing cognitive load. While advanced label aggregation techniques such as majority voting or confidence-based methods exist for classification tasks[15,29], they operate independently of annotator interactions and therefore fall outside the scope of HCI. Our focus is to apply models in data annotation and develop interactive pre-annotation tools that enhance the annotator user experience.

*3.3.2 Real-time LLM assistant*

A Real-time LLM assistant generates instant responses to user queries during annotation, making it particularly valuable for open-ended and creative annotation tasks. For example, in text response generation, the LLM produces an initial text response that the annotator can then revise or modify as needed. For record-keeping purposes and subsequent data integration, both the model-generated and human-modified responses are saved.

*3.3.3 LLM Judge*

The judge's role extends beyond reviewing subjects and labeling instructions; it also involves integrating quality rubrics into the prompt (illustrated Figure A1). We provide prompt templates and examples for point deduction and grading scale rubrics in Table A1 and A2, respectively. The judge evaluates annotations from human annotators or occasionally other models (e.g., LLM labelers), providing QA feedback per annotation, which includes a list of errors or grades for each criterion. QA feedback helps determine whether the labeled data meets the specified quality thresholds or requires further refinement. QA scores are automatically calculated using Eq. 1 or 2. As mentioned earlier, annotations marked as "REDO" are returned to the original annotators for revision and excluded from the download pipeline, while those marked as "PASSED" are considered "high quality" labels. Within the labeling UI, QA feedback tags are used to indicate whether the feedback was generated by the LLM Judge, auditors, or researchers.

## 4 EXPERIMENTAL RESULTS

### 4.1 Pre-annotation LLM Assistant for Comment Classification

*4.1.1 Problem statement*

Our goal was to explore the potential benefits of utilizing LLM-generated pre-annotations and confidence scores in classification tasks. Specifically, we aimed to improve annotation efficiency by reducing the average handling time (AHT) per annotation instance, while quantifying biases and overreliance on the LLM assistant's suggestions among human annotators.

**Annotation task**. We conducted an experiment using a dataset of comments from social media posts, which included a mix of images, videos, and text. Human annotators evaluated the relevance of comments to the post and their characteristics, providing a nuanced understanding of the data. We selected this dataset due to its multimodal nature and the subjectivity of the content, which requires human interpretation and confirmation, making it an ideal candidate for assessing the effectiveness of the LLM assistant in real-world scenarios. The annotation task consisted of three classification questions. The first question assessed the relevance of each comment to the post ("relevancy"), with the potential options being fully relevant, somewhat relevant, or not at all relevant. The second question assigned up to six characteristics to each comment including: informative or insightful, funny or humorous, emotional support, meaningful question, professional admiration or support, and self-disclosure. The final question determined whether the comment was civil and respectful. Our study focused exclusively on English-language posts and comments.

**Labeling UI.** As illustrated in Figure A2, the UI consists of two panels. The left panel displays the post and its corresponding comments, while the right panel presents the labeling options in a radio-button format for single-select questions or checkboxes for multi-select questions. The UI design aimed to reduce AHT per job by minimizing clicks and cognitive load. We achieved this through a simplified layout, grouping related questions and information together in a single page.

**Human annotation**. Experienced annotators from third-party vendors performed the annotation after undergoing training based on the labeling guideline for comment classification, which included instructions and examples for the three classification questions.

**Model**. We fine-tuned a Llama 7B model on human-annotated comment-post labels. For simplicity, we converted relevance and multi-select characteristic questions into binary "Yes" or "No" questions, resulting in 8 binary selection questions. To address class imbalance, we downsampled the majority class ("Yes") to match the minority class ("No") size. At inference time, the model was prompted with the post text, comment text, and a question posed at the end (Table 1). We iterated on the prompt, finding that a proper question and formatting guideline were essential for optimal performance, while including the definition of the characteristic could actually degrade performance. We observed high

AUC metrics (Figure 2), ranging from 0.84 to 0.99, when jointly fine-tuning all 8 questions together. The relevancy question proved more challenging than others, requiring more context related to multimodal information and understanding connections between posts and comments. In contrast, other questions focused more on comment texts themselves. The model's performance was deemed suitable for use as an LLM assistant. We also ensured no overlap between the fine-tuning data and the auditing data used for testing.

Table 1: Pre-annotation LLM assistant prompt template for comment classification

| Input | Template Section | Example |
|---|---|---|
| Post | ### Post:<br><br>{post} | ### Post:<br><br>Merry Christmas to All!!! |
| Comment | ### Comment:<br><br>{comment} | ### Comment:<br><br>Merry Christmas to you too Bobby! Hope you have a good one this year. Lots of love |
| Question | ### Question:<br><br>{question} | ### Question:<br><br>Does this comment ask meaningful questions? Answer Yes or No. |

**LLM assistant.** As a first step, the model generates confidence scores for each question, which were displayed alongside the labels to provide annotators with an indication of the model's certainty in its predictions (example UI in Figure 3). Labels with high confidence scores (greater than or equal to 0.5) were pre-selected. The goal was to provide annotators with a starting point for their evaluation, while also allowing them to adjust the labels and confidence scores as needed.

*4.1.2 Experiment Setup*

We established two queues for this experiment: a non-LLM assisted queue (Queue A) and an LLM-assisted queue (Queue B), which were created from previously audited jobs (illustrated in Figure A3). Both queues consisted of the same set of previously audited jobs. Each job presented a pair of comments and posts for annotation. We randomly split, 46 annotators to two equal-sized groups, with one group working on Queue A and the other on Queue B. The auditor labels served as the ground truth for evaluating label accuracy. Annotators has the options of rejecting jobs that cannot be reviewed due to language or tooling issues. After excluding those jobs, 1187 jobs remained for analysis.

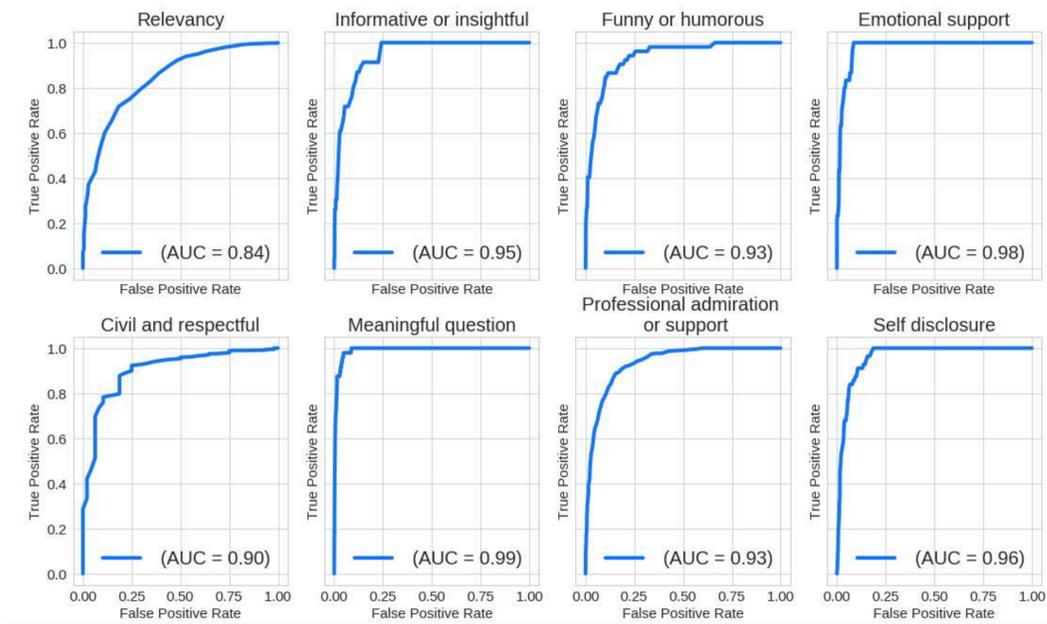

Figure 2: ROC Curves and AUC metrics for a fine-tuned Llama model used in the pre-annotation LLM assistant

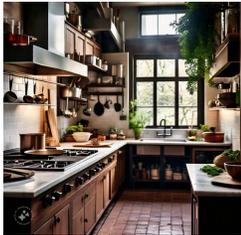

Figure 3: Example labeling UI for comment classification with pre-annotation LLM assistant, displaying pre-selection and confidence scores. The left panel shows a sample image (generated by Emu[8]) and comment for illustration purposes. The questions and answer options are consistent across all comment classification jobs

*4.1.3 Efficiency Analysis*

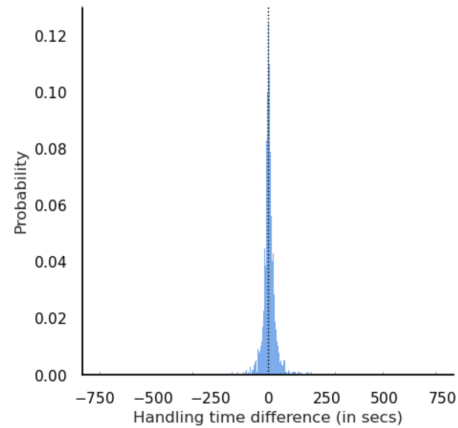

Figure 4: Handling time differences per job (s) with and without the pre-annotation LLM assistant

The distribution of the difference in handling times is shown in Figure 4. A one-sided paired t-test was used to compare the mean handling times between the two queues on 1187 jobs. The results showed that the handling time of a job from the non-LLM assisted queue was significantly longer than that of the LLM-assisted queue (p-value 0.0073), indicating that the LLM assistant had a statistically significant effect on reducing handling time. The estimated average handling time difference was 3.17 seconds, an average reduction of 12%, with a 95% confidence interval (0.63, 5.71).

*4.1.4 Human-LLM Agreement*

We also examined the relationship between handling time and human-LLM agreement. Our findings suggest that lower agreement rates between humans and LLM assistants generally result in longer review times. This is intuitive as annotators may spend more time deliberating on the correct label or seeking additional information to support their decision. To measure agreement, we used a 0-8 scale based on the number of matched answers between human annotators and LLMs. Because the handling time was continuous and the consistency metric was ordinal with 9 levels, we used Spearman's rho to measure the correlation and found that the correlation was -0.37 with a p-value below $10^{-42}$. The correlation between handling time and human-LLM agreement is strongest for jobs with high levels of disagreement. This suggests that the LLM assistant's suggestions are most effective when they align with the human annotators' initial judgment, and that deviations from this alignment can lead to increased review time.

*4.1.5 Quality Analysis*

To compare the label quality of the two queues, accuracy, precision and recall was evaluated for each of the 8 binary questions in Table 2 based on over the 1,187 audited jobs. Note that we assume the audit labels are the ground truth. Quality metrics showed marginal improvements when LLM assistants were used. The accuracy increased from 0.963 to 0.965, precision from 0.930 to 0.936, and recall remained constant at 0.931.

Table 2: Accuracy, precision, and recall metrics across 8 classification questions for both non-LLM-assisted and LLM-assisted queues.

| Question | Accuracy without LLM assistant | Accuracy with LLM assistant | Precision without LLM assistant | Precision with LLM assistant | Recall without LLM assistant | Recall with LLM assistant |
|---|---|---|---|---|---|---|
| Relevancy | 0.894 | 0.910 | 0.913 | 0.940 | 0.973 | 0.963 |
| Civil and respectful | 0.988 | 0.985 | 0.989 | 0.989 | 0.999 | 0.996 |
| Informative or Insightful | 0.976 | 0.963 | 0.625 | 0.400 | 0.484 | 0.452 |
| Self-Disclosure | 0.979 | 0.981 | 0.625 | 0.667 | 0.385 | 0.462 |
| Emotional Support | 0.997 | 0.998 | 0.000 | 0.500 | 0.000 | 0.500 |
| Funny or humorous | 0.992 | 0.992 | 0.200 | 0.200 | 0.200 | 0.200 |
| Professional admiration/support | 0.908 | 0.916 | 0.747 | 0.730 | 0.489 | 0.604 |
| Meaningful question | 0.956 | 0.971 | 0.650 | 0.776 | 0.456 | 0.667 |
| Overall | 0.963 | 0.965 | 0.930 | 0.936 | 0.931 | 0.931 |

### 4.2 Real-time LLM Assistant for VQA Text Generation

*4.2.1 Problem Setup*

This experiment aimed to investigate the benefits of leveraging the LLM assistant, where the model generates annotations in real-time during the labeling process. We focused on open-ended text generation tasks and evaluated key metrics, including annotation quality (measured by "Helpfulness" and "Honesty" as defined by Quality Rubrics), efficiency (measured by AHT), and annotator user experience. Our hypothesis was that introducing LLM assistants would lead to improvements in all three areas.

**Annotation task.** We designed an experiment using Visual Question Answering (VQA) annotation. In this task, annotators are asked to compose both a question and answer based on a given image, with the key constraint that the question must be created spontaneously during annotation, rather than being pre-filled or selected from a list. This approach enables us to capture the variability of human inquiry beyond pre-defined questions or prompts and simulate conversational dynamics. The end goal is to train models that can understand the visual context and accurately respond to a wide range of user questions about images. We selected VQA annotation also due to its open-ended and creative nature, which involves subjective overall quality goals including "Helpfulness" and "Honesty," each contains specific quality criteria (Table 3). During QA, auditors evaluate responses against each quality criterion, assigning a grade ("Good", "Minimum", or "Insufficient"). The grading scale rubrics, including detailed definitions for each grade, are outlined in Table A3. These annotations were then utilized as a part of SFT dataset for multimodal Llama 3[40] fine-tuning.

Table 3: Quality criteria under "Helpfulness" and "Honesty" for VQA annotations

| | |
|---|---|
| Helpfulness | Instructions Following |
| | Relevance |
| | Comprehensiveness |
| | Refusal |
| | Grammar and Presentation |
| | Language Consistency |
| | Tone / Style |

|  Honesty  |  Accuracy  |

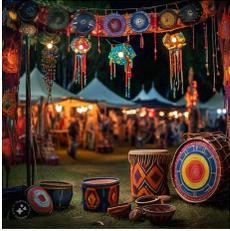

Figure 5: Side-by-side comparison of Annotation UI (a) without and (b) with real-time LLM Assistant for VQA annotations. Images (generated by Emu[8]) and text are provided for illustration purposes only.

**Human annotators.** We partnered with a third-party vendor to conduct the experiment. To ensure consistency and quality, we trained annotators with a comprehensive guideline, which included detailed instructions on utilizing the labeling UI and outlined specific quality criteria for annotations to be considered "Good".

**Labeling UI**. The UI (Figure 5a) features a two-panel design, with the left panel dedicated to query and response input. Two textboxes prompt annotators to enter a query and provide a corresponding response, while concise instructions and "Good" criteria definitions are displayed below, serving as quick reminders to guide annotators in crafting high-quality responses. The right panel displays the relevant image, along with additional contextual information such as the image caption.

**Model.** The model used was a checkpoint of Llama 3 87B multimodal model [40], which we believed could provide the baseline results in terms of answering visual questions. As is often the case, MILO leverages the model under development to refine and improve data annotation for subsequent training iterations. No prompt engineering was performed in this case, as we only passed in annotator query as the prompt.

**LLM assistant**. Between the query and response textboxes, an AI-powered textbox is provided with a generation button that produces LLM responses in real-time based on the annotator's input query (Figure 5b). When clicked, the button populates the textbox with the generated LLM response. To encourage creativity and reduce overreliance on LLM, we instructed annotators to use the generated response as a starting point and provide their own answer in the adjacent response textbox. We also instruct the annotators not to directly copy and paste LLM responses. This rule was also reinforced through both labeling guidelines and annotator training.

*4.2.2 Experimental setup*

We randomly sampled 714 images from previous VQA tasks and created two identical sets of image-question pairs. Two groups of 10 annotators worked on these pairs, one with the LLM assistant and the other without. To prevent any potential biases from memorized responses, we ensured that no annotator had previously worked on the same images. For this experiment, we employed a controlled setup by prefilling the questions across all annotation jobs and only allowing raters to provide answers (example UI in Figure A4). This approach enabled a direct comparison of the impact of LLM assistance on annotator responses. Note that this differs from our production setting, where annotators generate questions in real-time. By controlling for query difficulty, we aimed to isolate the effect of LLM assistance and ensure a fair comparison. The questions used in this experiment spanned multiple VQA categories, including Knowledge/Information Seeking, Scene Understanding, Text Understanding, Suggestion/Recommendation, Expression/Creativity, and Object Detection/Recognition.

*4.2.3 Quality Analysis*

Table 4: Average Word Count for annotator responses with LLM Assistant vs. without LLM Assistant

| Question category | Question Count | Annotator Response with LLM Assistant | Annotator Response without LLM Assistant |
|---|---|---|---|
| Knowledge / Information Seeking | 494 | 48.5 | 38.3 |
| Expression / Creativity | 45 | 42.6 | 42.7 |
| Text Understanding | 45 | 43.3 | 26.7 |
| Object Detection / Recognition | 44 | 58.8 | 32.7 |
| Scene Understanding | 40 | 90.1 | 48.9 |
| Suggestion / Recommendation | 46 | 78.5 | 35.4 |
| Weighted Average | | 52.70 | 37.91 |

As shown in Table 4, responses generated using the LLM assistant were 39.01% longer on average (52.70 words vs. 37.91 words) than those without it. The only exception was the Expression/Creativity category, where responses were only 0.1 words shorter. In contrast, responses in other categories saw length increases with average gains of at least 10.2 words. The most significant gains were seen in Suggestion/Recommendation (121% longer, +43.1 words), Scene Understanding (84% longer, +41.2), and Object Detection/Recognition (79% longer, +32.7 words). While longer responses may suggest more detail, word count alone is not a reliable metric of quality. LLM-assisted responses can contain redundant or unrelated information, and we observed instances where annotators included extraneous details that failed to address the question directly or incorporated hallucinated LLM responses into their answers.

Table 5: Head2Head Quality Evaluation for helpfulness and honesty for responses with LLM Assistant and without LLM Assistant.

| Question Category | Helpfulness Win Ratio (with LLM assistant) | Helpfulness Win Ratio (without LLM assistant) | Honesty Win Ratio (with LLM assistant) | Honesty Win Ratio (without LLM assistant) | Helpfulness Increase (with LLM assistant) | Honesty Increase (without LLM assistant) |
|---|---|---|---|---|---|---|
| Knowledge / Information Seeking | 24.7 | 24.7 | 17.4 | 13.6 | 0 | 3.8 |
| Expression / Creativity | 47.5 | 20 | 5 | 0 | 27.5 | 5 |
| Text Understanding | 31.1 | 11.1 | 31.1 | 13.3 | 20 | 17.8 |
| Object Detection / Recognition | 36.9 | 17.4 | 6.5 | 30.4 | 19.5 | -23.9 |
| Scene Understanding | 11.1 | 17.8 | 13.3 | 46.7 | -6.7 | -33.4 |
| Suggestion / Recommendation | 15.9 | 27.3 | 6.8 | 13.6 | -11.4 | -6.8 |
| Weighted Average | 25.96 | 22.88 | 15.90 | 15.61 | 3.09 | 0.28 |

We evaluated the LLM assistant's quality improvement using a blind, head-to-head comparison, where response sources were hidden in the labeling UI (Figure A5). A group of auditors evaluated the quality of responses in terms of helpfulness and honesty in a separate QA task, using predefined quality rubrics in Table A3. We also performed internal QA where the researchers evaluated the auditor responses, revealing an accuracy rate exceeding 90%. Such high level of accuracy demonstrates that the auditors' understanding of the labeling guidelines and quality criteria is comparable to our own, providing confidence in their assessments.

The evaluation focused on the "win ratio" – the proportion of times one response was deemed better than the other in terms of helpfulness and honesty. We analyzed this metric across the same query categories used in the word length analysis above. To ensure meaningful results, Table 5 presents the win ratios for each category, highlighting the frequency with which one response outperformed the other. For simplicity, we excluded instances where the responses were tie in terms of honesty or helpfulness. As shown in Table 5, the LLM assistant was effective in improving annotation quality, yielding a weighted average improvement of 3.09% in helpfulness and 0.28% in honesty across all categories. Notably, it achieved significant gains in helpfulness within the Expression/Creativity, Text Understanding, and Object Detection/Recognition question categories, as well as honesty in Text Understanding, with improvements exceeding 15%.

However, we observed some regression in honesty across certain categories, including Object Detection, Scene Understanding, and Suggestion/Recommendation. The LLM assistant also failed to provide substantial improvements in Scene Understanding and Suggestion/Recommendation for both helpfulness and honesty. These question types posed challenges for human annotators as well, and the model struggled to provide accurate answers. We found that LLM-assisted responses tended to include more details, but these additional details often failed to directly address the question and sometimes incorporated hallucinations (see Table A4 for examples). Consequently, the incorporation of these hallucinations into annotators' responses led to a notable regression in honesty. Furthermore, we noticed a substantial increase in response word length for these question types. This raised concerns about the efficacy of word length as a quality metric, as increased verbosity may have actually decreased the quality. To mitigate these issues, we recommended providing more explicit guidance to annotators on response style, emphasizing the importance of direct answers without unnecessary details. Additionally, guidelines should have enforced annotators to scrutinize the factual aspects of LLM-generated responses.

*4.2.4 Efficiency Analysis*

Table 6: Average handling time (AHT) for VQA annotation with and without LLM assistant

| Question | Handling Time with LLM Assistance (s) | Handling Time without LLM Assistance (s) |
| --- | --- | --- |
| Knowledge / Information Seeking | 444 | 389 |
| Expression / Creativity | 335 | 484 |
| Text Understanding | 289 | 156 |
| Object Detection / Recognition | 443 | 213 |
| Scene Understanding | 551 | 244 |
| Suggestion / Recommendation | 490 | 268 |
| Weighted Average | 436.26 | 353.54 |

Contrary to our hypothesis, integrating LLM assistants and response rewrites resulted in an unexpected increase in AHT. Specifically, responses generated with LLM assistants had a weighted average AHT of 436.24 seconds, which is 23.4% higher than the 354.54 seconds observed for responses without LLM assistance. This AHT increase raises questions about the trade-off between annotation quality gains and efficiency losses, highlighting the need to reassess the pros and cons of LLM assistants.

Notably, a strong correlation exists between the word count of LLM-generated responses and AHT, as illustrated in Table 6. First, LLM response generation can introduces a latency of a few seconds at inference time. Second, annotators need to spend more time to reading and digesting the typically long LLM responses, which in turn encourages them to develop more detailed and comprehensive responses. In fact, LLM-generated responses often consisted of 3-5 paragraphs, requiring annotators to invest significant time in reviewing and verifying accuracy while developing improved responses. In contrast, annotators without LLM assistance could immediately begin drafting their response to the query. It is worth noting that this was the annotators' first experience using LLM assistants for annotations. As they learned to effectively leverage the tool to enhance their responses, they inevitably incurred additional time costs, which also contributed to the increased AHT.

*4.2.5 Annotator User Experience*

We gathered annotator feedback from the vendor. In general, the feedback suggests that LLM assistants significantly mitigated creation fatigue, a common challenge for professional annotators who author numerous jobs daily, particularly when authoring creative and open-ended annotations. By providing a useful starting point, LLM-generated responses helped annotators overcome the blank slate problem, resulting in more detailed and structured answers. Moreover, annotators specifically noted that the framework proved particularly useful for identifying specific entities like plants and landmarks, highlighting the importance of targeted support in annotation tasks. In addition, LLM assistance also provides different aspects they did not cover. This finding underscores the design value of scaffolding, which reduces cognitive load and promotes user engagement.

### 4.3 LLM Judge for VQA Text Generation

*4.3.1 Problem Setup*

The traditional QA process, done by human auditors, allows annotators to refine their annotations based on feedback and enables researchers to filter out low-quality annotations, incorporating only high-quality labels in downstream model training. This experiment explored the potential of automating this process with the LLM judge, which evaluates human responses against predefined quality rubrics. By leveraging LLM Judge, teams can maintain high label quality while significantly reducing auditor workload, thereby expanding audit coverage and overcoming the manual nature of auditing. To evaluate the judge's performance, we measured its agreement with expert researcher feedback.

**Annotation task.** The original annotator responses were generated in the VQA task, as described in Section 4.2. As an auditor or researcher, their role is to judge the annotator responses by referring to the quality rubrics and selecting a grade for each criterion.

**Labeling UI.** The labeling UI displays the original annotator's view, including all context and question/options, as well as their responses. When the "Start QA" button is clicked, a form prompts the auditor to provide QA feedback on the annotator's responses based on each quality criterion from the rubrics, assign a categorical grade, and add explanations if needed. The QA scores are computed automatically, and the annotation is marked as either "REDO" or "PASSED" based on the threshold set for that annotation task, with feedback attributed to either "Auditor" or "Researcher". In annotators' review, the feedback is attached directly under the annotator response card, allowing them to filter by their own name and access the feedback by clicking "View Feedback" (Figure 6).

**Human Annotation**: The QA on annotator responses were performed by experienced researchers in this experiments. In general, we also reply on auditors from third-party vendors.

**Model**: We use Llama 3 70B with prompt engineering as the LLM judge. The model is prompted with inputs include project description, labeling instruction, review subject, annotator responses and definition for each quality criterion/grade. To accommodate lengthy quality rubrics with more than 10 criteria, we employ an iterative prompting approach. The LLM is prompted separately for each quality criterion, allowing it to select a specific grade. The prompt template is illustrated in Figure A1, with examples provided in Table A1 and A2.

**LLM Judge**: The judge evaluates quality based on grading scale rubrics to account for the subjective nature of the VQA tasks. In annotators' view, the judge's response is clearly labeled with the model's name and "LLM QA" (Figure 6). When expanding the "View Feedback" card, the judge's feedback is displayed for each quality criterion, accompanied by an explanation for the assigned grade, as seen in Figure 6.

*4.3.2 Experiment Setup*

We recruited 17 researchers with experience in AI data annotation and prior collaboration with the vendors. They were also familiar with the general quality requirements for annotation used to develop the Llama models. To ensure consistency and accuracy in their evaluations, we conducted a workshop before the labeling session. The workshop covered two key components: (1) quality rubrics outlining the criteria (Table A3) for assessing annotator responses, and (2) the format and content of LLM judge feedback as displayed in the UI. To reduce their workload, we focused on four key quality criteria from the "helpfulness" dimension: Comprehensive, Grammar & Presentation, Instruction-Following, and Tone/Style. These criteria are most critical in judging annotator responses and ensuring downstream model performance. Each researcher reviewed 10-50 pairs of annotator responses and corresponding LLM judge feedback. For each pair, they were asked to: (1) indicate whether they agreed or disagreed with the LLM judge's feedback, (2) provide alternative QA feedback with their own grade if they disagreed, and (3) focus on the assigned grade when indicating agreement, rather than the specific explanations provided by the LLM judge.

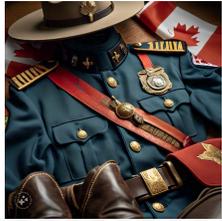

Figure 6: QA UI with LLM judge and human feedback. (a) Human auditor and LLM judge feedback examples. (b) LLM judge feedback example. The feedback assigns grades for each quality criterion, generating an overall QA score and determining "PASSED" or "REDO" status.

### 4.3.3 Quality Analysis

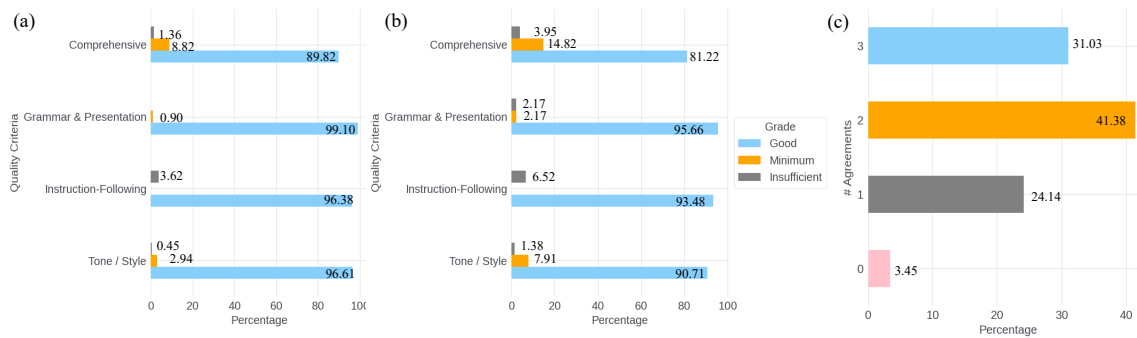

Figure 7: Comparison of LLM judge and human researcher feedback. a) Distribution of LLM judge feedback; b) distribution of human researcher feedback; c) percentage of grade agreements between LLM judge and human researcher within jobs where they have disagreed on at least one quality criterion

Table 7: Agreement rates between LLM judge and human researcher

| Quality Criteria in "Helpfulness" | Agreement rate % |
| --- | --- |
| Comprehensive | 83.08 |
| Grammar & Presentation | 91.91 |
| Instruction-Following | 91.91 |
| Tone / Style | 88.63 |
| Overall | 79.55 |

We randomly sampled 397 annotator responses, which were evaluated by both an LLM judge and human researchers. The results from the LLM judge feedback (Figure 7a) showed that the majority of responses received a "Good" grade across all four quality criteria, with the highest proportions of "Minimum" and "Insufficient" grades observed in the "Comprehensive" criterion at 8.82% and 1.36%, respectively. A similar distribution was observed in the human researcher feedback (Figure 7b), although they assigned slightly more "Minimum" and "Insufficient" grades across all four quality criteria. A detailed breakdown of agreement rates between the LLM judge and human researchers is provided in Table 7. We found an overall agreement rate of 79.55%, where both evaluators assigned identical grades for all four quality criteria. In cases where the LLM judge and human researchers disagreed, there was still significant concurrence on certain quality grades, resulting in higher agreement rates for each individual criterion. Specifically, high agreement rates around 90% were observed for Grammar & Presentation, Instruction-Following, and Tone/Style. In contrast, "Comprehensiveness" had the lowest agreement rate at 83.03%, as it is more nuanced even for human. Furthermore, analysis of the disagreeing cases (Figure 7c) revealed that over 31.03% had three agreeing grades (one disagreeing grade), while only 3.45% had no agreement at all (an example in Table A5). This suggests that even when the LLM judge and human researchers did not fully agree, there was often still substantial overlap in their evaluations.

Our previous experiences suggest that LLM judge performance is highly sensitive to the wording of quality criteria and error category definitions. Optimizing these definitions led to performance gains, including 11% higher precision and 6% higher recall. Effective LLM judge requires exact matching between review subject field names (e.g., "query", "annotator response") and those mentioned in rubric definitions (Table A3).

*4.3.4 Researcher user experiences*

The feedback gathered from researchers highlighted several areas for improvement. Many researchers felt that the judge's grades were often too harsh and would like to have the option to adjust the level of harshness to their preference. This flexibility allows them to balance between precision and recall, depending on the situation. To achieve this, we suggest incorporating adjustable harshness levels into the definition of each quality criterion and making it part of the LLM judge prompt. High recall is particularly important in this case, as it ensures that all problematic annotator responses are excluded from downstream training. However, it's also important to avoid being overly harsh, which could demotivate annotators. Additionally, researchers frequently reported that judge explanations were excessively verbose. To address this issue, we suggest adapting the explanation length to the annotator's response: concise explanations for brief responses and more detailed explanations for longer ones. This approach aligns with the principle of providing the right amount of information at the right time, enhancing the overall user experience. Overall, we recommend introducing a subjective input scale that allows users to customize their preferred level of harshness for each quality rubric. Furthermore, adjusting the LLM judge explanation length accordingly can be achieved through prompt engineering or by implementing an output token size limit.

## 5 DISCUSSION

### 5.1 Effectives of MILO framework

Our study has demonstrated the multifaceted potential of LLMs in AI data annotation, highlighting both benefits and caveats. By leveraging models to support professional human annotators, we diverge from previous studies that primarily compare LLM performance to crowdworkers. In addition, we focus on multimodal annotation workflows, including both classification and text generation based on images, offering valuable insights into the practical applications of LLMs in real-world, production annotation scenarios and foundation model development.

**Enhanced annotation efficiency.** Our results show that the pre-annotation LLM assistant can support annotator decision-making and reduce annotation time by 12% using the fine-tuned Llama 7B model (Section 4.1). The real-time LLM assistant alleviates annotator creation fatigue and provides structures for text generation tasks. Furthermore, the LLM judge agrees with expert researchers 79.55% of the time, indicating its potential as an effective judge to expand QA coverages and optimize valuable auditing resources. These findings suggest that LLMs can streamline the annotation process and reduce human workload.

**Quality improvement.** The LLM-assisted pre-annotations maintain, or moderately increase accuracy compared to non-LLM assisted labels, suggesting that LLM assistants can enhance annotation quality, particularly when consistency across annotators is crucial. With real-time LLM assistance, we observe a 3.09% increase in helpfulness and a 0.28% increase in response honesty, with significant improvements in certain question categories such as Text Understanding, Expression/Creativity, and Object Detection/Recognition. Additionally, the LLM judge can filter out low-quality labels, boosting the quality of annotations used in downstream model training. It is also worth noting that improvements in quality may come at the cost of reduced efficiency, as evidenced by a 23.4% increase in AHT (Section 4.2). Therefore, researchers should operate at a trade-off point that best aligns with their specific annotation needs and goals.

**Model performance.** Model performance has a significant impact on efficiency and quality. Our results show that lower agreement rates between humans and LLM assistants lead to longer review times (Section 4.1). Building on the promising gains of a fine-tuned Llama 7B model, more advanced models such Llama 3.1 405B or GPT-4 can further improve accuracy for pre-annotations, reduce annotator disagreements, and enhance annotation efficiency and quality.

Conversely, poor model performance can have a negative impact, such as model hallucination affecting human annotation honesty (Section 4.2) and potential low recall in the LLM judge missing low-quality annotations (Section 4.3). Therefore, it is essential to properly benchmark model performance against gold standards or audit labels (if available), followed by detailed human evaluation (as outlined in Section 4.3), before deploying in production.

**Mitigating Biases in Human Annotation.** The interactive nature of MILO inevitably introduces biases into the human annotation process, as seen in Section 4.2 where annotators are influenced by model hallucination. However, these biases can be mitigated through better annotator education, including clear labeling guidelines and training sessions. Notably, annotators spend more time reviewing pre-annotations when they disagree with the model (Section 4.1), indicating that professional annotators critically evaluate model suggestions rather than blindly accepting them - a key distinction from crowdworkers who often prioritize speed and quantity over accuracy and quality. By incorporating an additional QA process, such as the LLM judge (Section 4.3), we can further ensure the accuracy, reliability, and consistency of annotations. As the paradigm shifts in human annotation, it is essential to prioritize quality over quantity, investing in robust QA processes and tooling that empower annotators to produce high-quality annotations that meet the rigorous standards defined by research teams.

## 5.2 Designing human-AI collaborative annotation system

When designing the MILO framework, we considered several key design principles for generative AI applications [49]. Our learnings from this process have implications for the design of human-AI collaborative systems. The interaction pattern between humans and AI in the MILO framework depends on the specific annotation task, such as classification or text generation. Additionally, the annotation workflow can be divided into three stages: pre-annotation, real-time assistance, and post-annotation QA. The stages and interaction pattern serve as the basis for how we design the three major applications and the experiments.

**Embracing Imperfection and Uncertainty.** We can design the MILO system to acknowledge the imperfections and uncertainties of AI-generated outputs. The LLM assistant and judge provide confidence scores and uncertainty indicators, helping annotators understand the AI system's limitations.

**Enabling Co-Creation**. The MILO framework is designed to facilitate co-creation between humans and AI, enabling annotators to actively participate in the generative process. For LLM assistant, this is achieved through various mechanisms, including adjusting pre-selections in classification tasks and sending input queries in text generation tasks. Additionally, annotators can edit and refine generated outputs. In the case of the LLM judge, researchers can provide their own input to the prompt, such as labeling instructions, project descriptions, and quality rubrics. Furthermore, reflecting user feedback, we can include more generic input parameters that allow users to control the generative process. These may include settings such as random seed, output word limit, levels of harshness, or other customizable options. By providing these controls, we can empower users to tailor the generative process to their specific needs and preferences, fostering a more collaborative and effective human-AI partnership. MILO's gentle guidance enables annotator to focus on higher-level thinking, producing high-quality annotations that align with human-centered design principles, emphasizing intuitive tools that support users' workflow and cognitive abilities.

**Building Appropriate Trust and Reliance.** To ensure that annotators use the MILO effectively, we designed it to provide explanations and rationales. For instance, the pre-annotation LLM assistant includes explanations for why it chose certain options. The LLM judge provides clear explanations of its decisions and rationales for outputs. This approach aligns with established guidelines for human-AI interaction [2]. However, additional mechanism is needed to enable the LLM to admit when it does not know the accurate answer. The LLM should provide upfront explanations of its capabilities and limitations, enabling annotators to make informed decisions about when to trust on its outputs, and when to question them. To mitigate overreliance and biases, we could also introduce subtle friction points in the workflow, such as banners or pop-ups cautioning that results may be inaccurate. This encourages annotators to review and evaluate outputs critically, ensuring they remain engaged and vigilant throughout the annotation process.

### 5.3 Limitations and future work

While the MILO framework demonstrates promising results, there are opportunities for improvement. Specifically, we relied on fine-tuned smaller models or prompt engineering, which may not have achieved optimal performance; more advanced techniques such as Retrieval-Augmented Generation (RAG) with few-shot examples[12,22] and Chain-of-Thought (CoT) prompting[25], or more sophisticated models, could potentially yield better results. Additionally, optimizing inference time to reduce latency of interactive LLM assistants can further reduce AHT and enable real-time feedback, potentially allowing for an online, interactive LLM judge that provides immediate feedback. Due to resource constraints, we did not implement a mechanism to collect annotator feedback within the labeling UI, instead relying on vendor interviews; future studies should prioritize developing more comprehensive feedback mechanisms that allow annotators to provide feedback and input into the generative process.

## 6 CONCLUSIONS

In this work, we present the MILO framework, a collaborative system that combines the strengths of professional human annotators and LLMs in AI data annotation tasks. We explore the effectiveness of LLMs as pre-annotation and real-time assistants, and judges on annotator responses, depending on the interactive pattern between annotators and models. For efficiency and operational benefits, LLM assistants improve labeling efficiency by reducing handling time and alleviating annotator creation fatigue. Additionally, they increase specific quality criteria such as "helpfulness," which is critical for downstream LLM fine-tuning. LLM judges provide feedback that aligns with human auditors' judgments, enabling teams to scale up the QA process and optimize auditing resources. We also introduce quality rubrics that define quality criteria for open-ended, creative annotations, which are often difficult to quantify. These rubrics are flexible enough to handle both error detection and grading scale modes, prioritizing accuracy or providing fine-grained feedback, respectively. Overall, our results show that integrating models into the annotation loop enhances labeling efficiency, benefiting human annotators, while also helping ML teams reduce their reliance on human annotation alone and improving data annotation quality. Such efforts, in turn, accelerates AI/ML model development and ensures better alignment between human and machine values.

**ACKNOWLEDGMENTS**

We acknowledge Arup Kanjilal, Ryan Owens, Anil Tripathi, Sung Park, Shuo Tang, Ai-jung Li, Rishi Godugu, Thai Quach, Edward Tseng, Xin Peng, Hrish Mula, Rena Liang, Eric Alamillo, Lawrence Ma, Oluseun John, Lydia Smith, Enoch Chen, EvaRose Dwyer, Beth Huxster, Anthony Yun, Chien Yu, Joshua Celli, Saurabh Verma, Viet-An Nguyen, Yilin Zhang, Annie Dong, Pratik Dubal, Kun Huang, Jules Bozhich, Kaiea Rohlehr, Fatima Caballero, Liz Belile, James Qian for their valuable contributions to the MILO framework and dedicated efforts in data annotation.

# A APPENDICES

## A.1 LLM judge prompt template

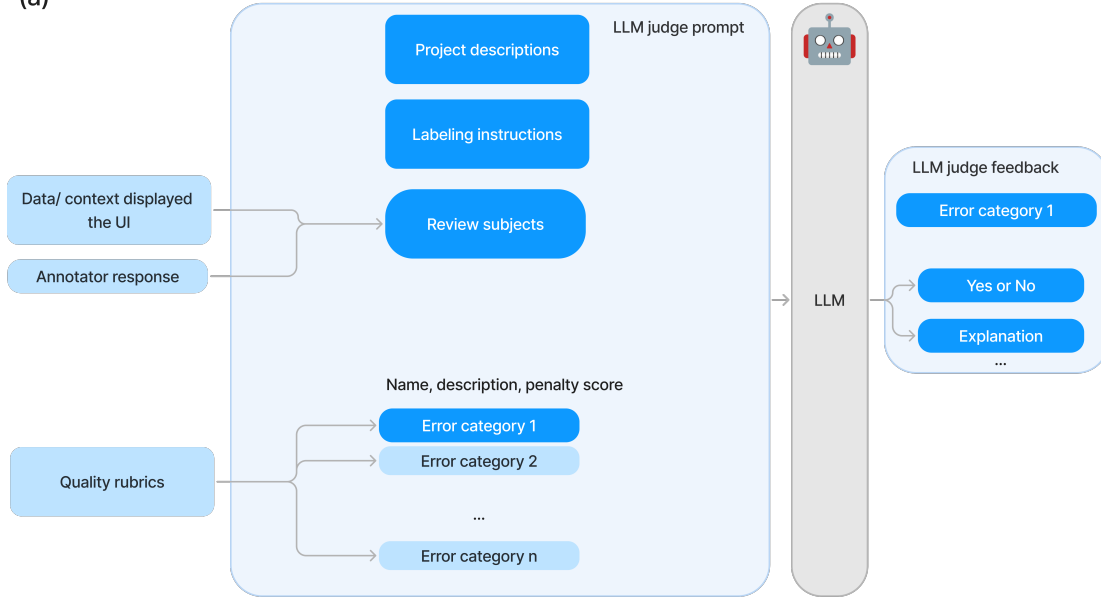

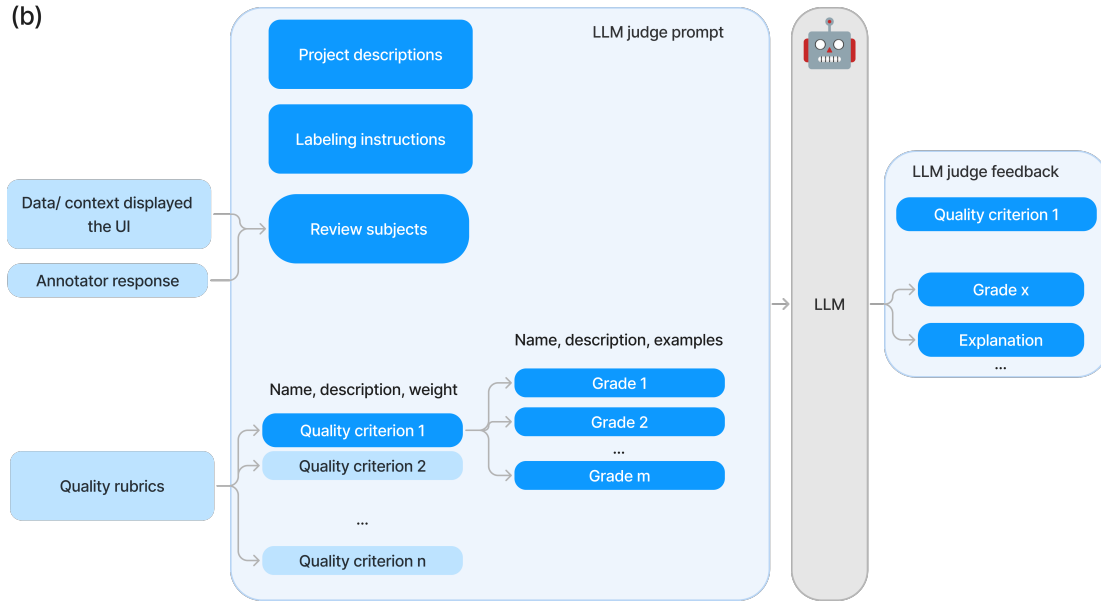

Figure A1: Illustration of LLM judge prompt template for a) point deduction rubrics b) grading scale rubrics

Table A1: Prompt Template for LLM Judge Point Deduction Rubric

| Template Section | Template with input variables in {} | Example |
|---|---|---|
| Introduction | *As a quality auditor within the quality control team, you have been provided with several key documents and guidelines to aid the review process. Your primary task is to evaluate the {qa_field_of_interest} based on the provided materials and identify any instances of {error_category_name} errors, as defined below.* | *As a quality auditor within the quality control team, you have been provided with several key documents and guidelines to aid the review process. Your primary task is to evaluate the annotator_response based on the provided materials and identify any instances of Answer does not meet Good criteria errors, as defined below.* |
| Project Descriptions | *{project_description}* | *The project's goal is to enhance Llama Model Performance for Image-based Visual Question Answering through Supervised Fine-Tuning and Annotation.*<br><br>*The annotators are provided with a query related to an image and responses to the query. Your task is to evaluate the quality of the annotator response based on predefined quality guideline.*<br><br>*The queries should resemble questions users would ask an AI assistant. Initially, we generate a model response for annotators' reference, followed by annotators providing their own responses adhering to the 'Good' criteria.* |
| Labeling Instructions | *{labeling_instructions}* | *Annotators are given the following instructions:*<br>*[START ANNOTATOR INSTRUCTIONS]*<br>*When providing query:*<br>*Imagine you are a local user of AI assistant, please make sure your query is related to the concept of that country (see Context field).*<br><br>*When providing response:*<br><br>*Considering the following aspects meeting the good criteria*<br>*Accuracy - The extent to which the information presented is accurate, reliable, and aligns with established facts or evidence.*<br>*Relevance - How useful the supporting information and claims are in answering the question or prompt.*<br>*Safety - The written response should not contain any dangerous/illegal content, unqualified advice and compromise youth safety.*<br>*Grammar and Presentation - The distinctive method in which ideas are expressed through writing focusing primarily on the stylistic, mechanical and syntactical components.*<br>*Instructions Following - The extent to which the answer addresses all aspects of the prompt, ensuring that no essential information is omitted.*<br>*Tone / Style - The overall writing style or "voice" of the response, which should be consistent within and between responses.* |

| Template Section | Template with input variables in {} | Example |
|---|---|---|
| | | *Signal to noise - The response does not convey unnecessary or superfluous information. It is as "snappy" as it can be, while still fully and comprehensively answering every component of the prompt.*<br>*Depth - The level of detail, insight, and nuance the response provides.*<br>*Delightfulness - The degree to which the response sparks joy and delight in the user when applicable.*<br>*Humor - The response uses humor effectively and appropriately when applicable.*<br>*Cultural Context - The response appropriately captures and incorporates the cultural context of the prompt or topic.*<br>*Above and Beyond - Does the model go beyond the prompt that was not explicitly requested by the user, but would be directly helpful and relevant to do so.*<br>*And your response is better than model response on the above aspects.*<br>*[END ANNOTATOR INSTRUCTIONS]* |
| Review Subjects | *{review_subjects}* | *"model_response": "The plant in the image is an Astrophytum asterias, also known as a star cactus. It is a small, globular cactus that typically grows to be around 2-5 inches (5-13 cm) in diameter. However, some specimens can reach up to 8 inches (20 cm) in diameter or more in ideal conditions.\n\nAstrophytum asterias is a slow-growing plant, and it can take many years for it to reach its full size. Factors such as light, temperature, watering, and fertilization can all impact the growth rate of this plant. In general, it is recommended to provide Astrophytum asterias with bright, indirect light, moderate temperatures (around 65-75\u00b0F/18-24\u00b0C), and infrequent watering. Fertilization should be done sparingly, as over-fertilization can damage the plant.", "annotator_response": "The plant in the image appears to be an Astrophytum asterias or star cactus. It is a small, round cactus that grows to a height of 2.5\u20136 cm (1\u20132 in) and a diameter of 5\u201315 cm (2\u20136 in).", "Query ": "How big do these plants normally grow?", "Caption": "Caption: astro background flower image "* |
| Error Category Name | *{error_category_name}* | *Answer does not meet Good criteria* |
| Error Category Definition | *{error_category_definition}* | *Answer does not meet the good criteria on the guideline* |
| Conclusion | *After your assessment, please respond with 'Yes' if you find any errors of the specified category, or 'No' if no such errors* | *After your assessment, please respond with 'Yes' if you find any errors of the specified category, or 'No' if no such errors are present. Additionally, provide a brief explanation for your* |

| Template Section | Template with input variables in {} | Example |
|---|---|---|
| | *are present. Additionally, provide a brief explanation for your decision, referencing specific aspects of the guidelines or the review subjects as necessary.* | *decision, referencing specific aspects of the guidelines or the review subjects as necessary.* |

Table A2: Prompt template for LLM judge grading scale rubrics

| Template Section | Template with input variables in {} | Example |
|---|---|---|
| Introduction | *As a quality auditor within the quality control team, you have been provided with several key documents and guidelines to aid the review process. Your primary task is to evaluate the {qa_field_of_interest} based on the provided materials and provide your grading scale for {error_category_name} quality criteria. Please be a fair and unbiased judge, do not be overly critical or forgiving. The scale and the respective definition is defined below:*<br>*{levels and definitions}* | *As a quality auditor within the quality control team, you have been provided with several key documents and guidelines to aid the review process.*<br>*Your primary task is to evaluate the annotator_response based on the provided materials and provide your grading scale for Cultural Context quality criteria.*<br>*Please be a fair and unbiased judge, do not be overly critical or forgiving.*<br>*The scale and the respective definition is defined below:*<br>*- \*\*N/A\*\*: Not appliable*<br>*- \*\*Minimum\*\*: The response shows a basic understanding of the cultural context of the prompt.\nIt includes some relevant cultural references or nuances, but may not fully capture the depth of the culture*<br>*- \*\*Good\*\*: The response shows a deep understanding and incorporation of the cultural context of the prompt. It appropriately includes cultural nuances and provides a culturally relevant response.\nThe cultural context enhances the overall quality of the response.*<br>*- \*\*Insufficient\*\*: The response shows a lack of understanding or incorporation of the cultural context of the prompt.\nIt may miss or ignore important cultural nuances.* |
| Project Description | *{project_description}* | *The project's goal is to enhance Llama Model Performance for Image-based Visual Question Answering through Supervised Fine-Tuning and Annotation.*<br>*The annotators are provided with a query related to an image and responses to the query. Your task is to evaluate the quality of the annotator response based on predefined quality guideline.*<br>*The queries should resemble questions users would ask an AI assistant. Initially, we generate a model response for annotators' reference, followed by annotators providing their own responses adhering to the 'Good' criteria.* |
| Labeling Instructions | *{labeling_instructions}* | *Annotators are given the following instructions:*<br>*[START ANNOTATOR INSTRUCTIONS]* |

| Template Section | Template with input variables in {} | Example |
|---|---|---|
| | | *When providing query:* *Imagine you are a local user of AI assistant, please make sure your query is related to the concept of that country (see Context field).* *[END ANNOTATOR INSTRUCTIONS]* |
| Review Subjects | {review_subjects} | *"query": "Is the Cape Town International Jazz Festival usually held at a outdoor venue like the concert stage in the picture?", "model_response": "No, the Cape Town International Jazz Festival is usually held at the Cape Town International Convention Centre (CTICC), which is an indoor venue. The festival features multiple stages, including the Kippies Stage, the Moses Molelekwa Stage, and the Rosies Stage, all of which are located inside the CTICC. The venue provides a controlled environment that allows for better sound quality and a more intimate experience for the audience. While outdoor concerts can be great, the Cape Town International Jazz Festival is typically held indoors to ensure a high-quality experience for both the performers and the attendees.", "annotator_response": "The Cape Town International Jazz Festival is usually not held at outdoor venues. Instead, the festival is generally held in the Cape Town International Convention Centre. Let me know if you're interested in learning more about this event!", "Context ": "Please provide VQA annotations about **Cape Town International Jazz Festival** in **South Africa**",* *"Caption": "Electric nights and unforgettable performances at the concert hall"* |
| Quality Criteria Name | {quality_criteria_name} | *Cultural Context* |
| Quality Criteria Definition | {quality_criteria_definition} | *The response appropriately captures and incorporates the cultural context of the prompt or topic.* |
| Grade Options | *After your assessment, please respond by choosing one of the following rating scale options: {list_of_grade_names} Additionally, provide a brief explanation for your decision, referencing specific aspects of the guidelines or the review subjects as necessary.* | *After your assessment, please respond by choosing one of the following rating scale options:* *- **N/A*** *- **Minimum*** *- **Good*** *- **Insufficient*** *Additionally, provide a brief explanation for your decision, referencing specific aspects of the guidelines or the review subjects as necessary.* |

## A.2 Comment classification

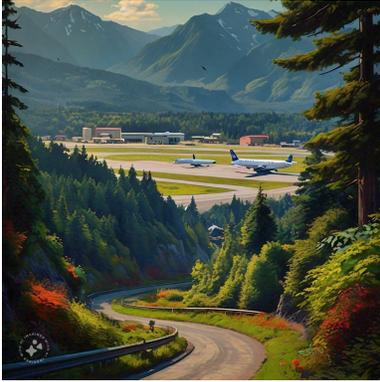

Figure A2. Example labeling UI for comment classification without LLM assistance. The left panel shows a sample image (generated by Emu) and comment for illustration purposes. The questions and answer options are consistent across all comment classification jobs.

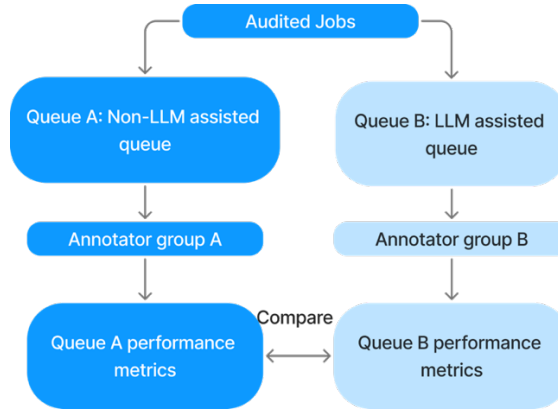

Figure A3: Experiment setup for Section 4.1, where the audited jobs are the same across both the non-LLM assisted and LLM assisted queues.

## A.3 VQA Text Generation Annotation

Table A3: Quality Rubric for VQA Text Generation Annotation used in Section 4.2 and 4.3

| Quality Criteria | Definition | Grade 1 (Insufficient) | Grade 2 (Minimum) | Grade 3 (Good) |
|---|---|---|---|---|
| Honest | Accuracy *When a verifiable claim is made, the extent to which the information presented is accurate, reliable, and aligns with established facts or evidence.* | Central Claims - One or more incorrect claims are central or core to the conclusion, thesis, or key supporting points of the response.<br>Supporting Claims - 2 or more pieces of supporting evidence are false<br>Verifiable - Conclusive, common-knowledge counterfactual information can be found via a Google search and/or falsely asserts claims or evidence that are not fully proven / controversial as fact | Central Claims - All central claims are correct and accurate. The central claim is the thesis statement of the response. It's the part of the response that directly answers the user's question.<br>Supporting Claims - Up to 1 incorrect supporting point that, if corrected, would not meaningfully alter the core thesis or conclusion. Supporting claims provide evidence or backup for the central claim.<br>Verifiable - Falsely asserts claims or evidence that are not fully proven / controversial as fact but does not meaningfully alter the core conclusion | Central Claims - All central claims are correct and accurate.<br>Supporting claims - All supporting claims are accurate and correct<br>Verifiable - No claims or evidence are incorrect or false, and are verifiable through general available information, and trusted sources. All potentially controversial, generalized statements or opinions are presented with appropriate caveats<br>Citation - Reputable sources are cited where appropriate to back up claims |
| Helpful / Content Quality | Instruction-Following *The extent to which the answer addresses all aspects of the prompt, ensuring that no essential information is omitted.* | Prompt Request Coverage - The response does not address all explicit asks of the prompt<br>Constraints - The response does not satisfy all parameters or constraints provided within the prompt. | | Prompt Request Coverage - The response addresses all explicit asks of the prompt<br>Constraints - The response satisfies all parameters or constraints provided within the prompt. |
| | Relevance *How useful the supporting information and claims are in answering the question or prompt.* | Supporting content - The key points provided are unrelated to the central claim or thesis<br>Supporting Context - Essential details are omitted, making the output difficult to understand in the context of the prompt.<br>Specificity - The examples brought up are vague and overly general<br>Usefulness - The information provided to justify or support any of the key points of the response | Supporting content - The supporting information and evidence is tangentially related to the central claim or thesis<br>Supporting Context - The response provides the key supporting ideas that make the response cohesive, but may omit or ignore minor background information<br>Specificity - The information provided contains a few selected examples, but doesn't explain and relate them back to the original response | Supporting content - The supporting information logically defends or clearly illustrates the key points, and the central claims made in the response<br>Supporting Context - The output is rich in context, with every idea well-supported and seamlessly integrated into the overall narrative.<br>Specificity - The claims are supported with precise, rigorously chosen, examples that are related back to the original claims<br>Usefulness - Each word, paragraph and sentence directly add value to the claims or context of the response. There is no filler or superfluous information. |

| | | does not logically contribute to the main claim | Usefulness - There is a mix of useful, and irrelevant but not directly useful information within the response. The core requests/demands of the prompt are still satisfied within the response, despite the irrelevant information | |
|---|---|---|---|---|
| | Comprehensiveness *The level of detail, insight, and nuance the response provides.* | Justification - There is no supporting information at all. The response does not include meaningfully relevant insights, evidence, details, or examples.<br>Context - The response omits important or necessary background information to understand the content of the response.<br>Logical Reasoning - No rationale is provided at all<br>New Concept Development - New ideas introduced are poorly connected or irrelevant to the premise of the response. The response may feel scattered and disorganized, with ideas that do not logically contribute to the overall topic. There is little to no engagement or cohesiveness, making the response difficult to follow and understand.<br>Comprehensive - The response is significantly lacking in detail and fails to cover essential aspects of the question or request. Information provided is minimal, superficial, or incorrect, leaving the user with an incomplete or misleading understanding. The user is likely to feel frustrated and unsatisfied due to the lack of valuable and pertinent information. | Justification - The supporting information was helpful in understanding and justifying the claims but was surface-level. There was analysis missing that is usually not found on a cursory google search.<br>Context - The response provides the most important or necessary background context, but may omit or ignore minor background information<br>Logical reasoning - The chain-of-thought-reasoning is included, but shallow, and can be more complete, and/or step-by-step.<br>New Concept Development - New ideas introduced in the response have some relevance to the premise but may lack seamless integration or fail to add significantly to the overall cohesiveness of the response. There may be some disjointedness or tangential information that does not fully support the main topic.<br>Comprehensive - The response provides a basic level of information but lacks thoroughness and detail. Some key aspects of the question or request may be only partially addressed or overlooked entirely. The user may come away with an incomplete understanding or feel that additional information is needed to fully satisfy their query. | Justification - The supporting information includes ample relevant insights, evidence, details, and examples. The supporting information and evidence are exhaustive in regards to supporting the set of claims (i.e., there is a specific example or justification for each main claim). Reputable sources are cited where appropriate to back up claims.<br>Context - The response provides the necessary useful background information and context for the user to understand the response<br>Logical reasoning - The chain-of-reasoning of the response is sufficient and addresses most, if not all, immediate follow-up/related questions a user may have<br>New Concept Development - All new ideas introduced connect smartly to the premise of the response and together they make for an engaging, and cohesive response.<br>Comprehensive - The response feels comprehensive. It provides thorough and detailed information that covers all aspects of a specific question or request, ensuring that the user feels fully informed and satisfied when coming away from a conversation. | |

| | | | | |
|---|---|---|---|---|
| | Refusal<br>*When applicable: if the response is a refusal, evaluate whether it is a true refusal.* | False Refusal - The response is a false refusal according to our product specs and safety guidelines. | | True Refusal - The response is a true refusal according to our product specs and safety guidelines. |
| Helpful / Language Quality | Grammar & Presentation<br>*The distinctive method in which ideas are expressed through writing focusing primarily on the stylistic, mechanical and syntactical components.* | Formatting - The way that the response is visually presented detracts from readers ability to understand the content. There is little or no visual separation between ideas. Lists are not broken into bullet points and Markdown is broken and unable to render properly.<br>Spelling/Grammar - The response has multiple spelling, punctuation, or grammatical errors that significantly impact how easily the response can be parsed by human readers. | Formatting - The visual presentation of the response does not impede the reader's ability to understand the response, but there are parts of the response that could be more richly formatted for improved readability/understandability.<br>Structure - The response uses sentence structure that matches the context of the prompt or question. Response avoids using generic sentence structure.<br>Spelling/Grammar - The response has some minor spelling or grammatical errors, but the response is still readable/understandable. | Formatting - The visual presentation of the response directly contributes to the readability/understandability of the response. Complex tools such as LaTex and markdown are used appropriately and when applicable to advance the comprehension of the response<br>Structure - The response uses sentence and paragraph structure that matches the context of the prompt or question. Response avoids using the same structure for every sentence and response.<br>Spelling/Grammar - The response has no spelling, punctuation, or grammatical errors. |
| | Tone / Style<br>*The overall writing style or "voice" of the response, which should be consistent within and between responses.* | Sentiment - Responses sound robotic, do not mirror the fluidity and subtleties of human conversation. Responses are not straightforward and contain mostly opinions. The response is either overly friendly or overly serious without consistency.<br>Informativeness - Responses contain mostly opinions. Responses are always incomplete.<br>Intelligence - Responses do not sound smart and straightforward, and do not directly address the prompt. Use of unnecessary jargon and technical terminology that does not reflect user input. Responses incorrectly assume intent from the user, and the response sometimes | Sentiment - Responses do not always sound natural and human-like, inconsistent in mirroring the fluidity and subtleties of human conversation. Responses are not straightforward and contain opinions and at times judgmental, the response can be overly friendly or overly serious without consistency.<br>Informativeness - Responses contain facts and opinions. Responses are not always comprehensive, do not answer the question completely and lacking in information that would enrich the response.<br>Intelligence - Responses do not consistently sound smart and straightforward, and do not directly address the prompt. Use of unnecessary jargon and technical terminology that does not reflect user input. Responses incorrectly | Sentiment - Responses sound natural and human-like, effectively mirroring the fluidity and subtleties of human conversation. Responses should be straightforward and non-judgmental, neither overly friendly nor overly serious.<br>Informativeness - Responses are centered on information rather than opinion or emotion. Responses should always feel comprehensive, answering the question asked with as much detail as is helpful, without editorializing.<br>Intelligence - Responses should sound smart and straightforward, directly addressing the prompt. Use of jargon and technical terminology should reflect user input. Responses should assume good intent from the user but should not come across as overly naive.<br>Engagement - Responses should ask questions if there are unanswered elements of the conversation, but follow-up questions should provide a specific purpose. Responses show empathy for the user's stated issue while still guiding them toward a solution. |

| | | comes across as overly naive or overly defensive. Engagement - Responses never ask follow-up questions if there are unanswered elements of the conversation. Responses do not show empathy for the user's stated issue when applicable and do not always guide them toward a solution. | assume intent from the user, and the response sometimes comes across as overly naive or overly defensive. Engagement - Responses sometimes ask questions if there are unanswered elements of the conversation, but follow-up questions do not always provide a specific purpose. Responses are inconsistent in showing empathy for the user's stated issue and do not always guide them toward a solution. | |
|---|---|---|---|---|

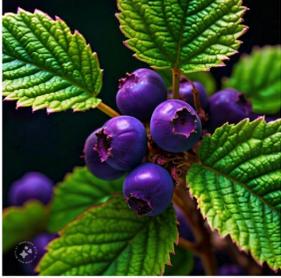

Figure A4: Side-by-side comparison of Annotation UI in VQA experiment (a) with and (b) without LLM assistant in section 4.2. Images (generated by Emu) and text are for illustration purposes.

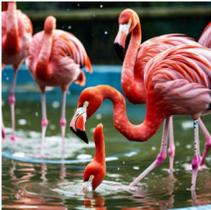

Figure A5: Head2Head Comparison UI to compare the quality of the two responses. We made it straightforward to distinguish and select between the questions above as well as indicate the winner of both helpfulness and honesty. The response sources are blind to the auditors. Images (generated by Emu) and text are for illustration purposes.

The following examples illustrate the impact of LLM assistance on annotator responses. We compare responses with and without LLM assistant, as well as the LLM output provided to the annotator. In example 1, the annotator response with LLM assistant includes detailed descriptions from the LLM that are not present in the image, resulting in an appropriate penalty. In contrast, the response without LLM assistant does not exhibit this issue. In example 2, while the response provides great detail about each object in the image, including entity counting, the annotator incorrectly indicates the quantity of chairs, compromising the factuality of the response. Notably, entity counting is less prevalent in responses without LLM assistant. Examples 3 and 4 demonstrate how LLM assistance can affect the honesty score. In both cases, the annotator confidently names the object using the LLM response, even when the image is difficult to discern according to the evaluation judge. However, this is due to the annotator's capability rather than the LLM response itself. Finally, example 5 shows how LLM assistant can introduce bias in terms of helpfulness. The annotator response focuses extensively on entity details, mirroring the LLM output, but fails to answer the question until the end of the response. This highlights the need for careful consideration of LLM assistant in annotator responses.

Table A4: Examples of underperforming cases using LLM assistant. Images (generated by emu) and text are for illustration purposes.

| # | Question category | Image | Question | Annotator response with LLM Assistant | Annotator response without LLM Assistant | LLM Response |
|---|---|---|---|---|---|---|
| 1 | Scene Understanding | 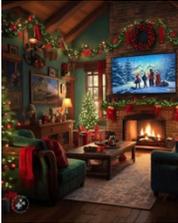 | Describe the scene. | The image displays a living room that has been decorated for Christmas. A television is bright upon the wall, placed between two shaded windows and above a wooden cabinet where autumnal decorations and a Sony PlayStation 5 are placed. Red, green, blue, pink and gold, colored lights beam from the Christmas tree stood in the corner and fill the room with a subtle, warm light. Drink tumblers, solo cups and open cans, signs of consumed beverages are placed across the matte black table in the foreground. The dark socks of a person's feet lean in from the right, nestled upon the black-and-white plaid pattern upon the couch, near a pink teddy bear laid across the floor. | This scene depicts a living room, dimly lit by the light from the television mounted on the wall and the Christmas tree with multicolored lights. A large cabinet rests against the wall below the TV, between two windows with blinds. Drink cans and cups sit on a table in the foreground. | The image shows a dimly lit living room decorated for Christmas. The Christmas tree, on the right side of the image, is lit up with multicolored lights and decorated with ornaments. There is a person sitting on the couch, facing away from the camera. They are wearing a plaid shirt and have dark hair. There is a TV mounted on the wall, showing a movie or TV show. The TV is in the middle of the image. There are two windows with blinds on either side of the TV. There is a dark wooden cabinet underneath the TV, with red and orange leaves and a few pumpkins on top. The floor is tiled and there are a few paper plates and red cups on the coffee table in front of the couch. The room is dimly lit, with only the lights from the TV and Christmas tree providing illumination. |
| 2 | Scene Understanding | 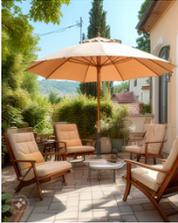 | What's the photo about? Tell me about the picture. | This picture captures a tranquil sitting area on sunny day. A laptop is open beside a green can of beverage and two books on a round table. A green umbrella with a wooden pole is positioned through the center of the table. Five plastic chairs are white and clean, positioned casually near the table. | This photo features a quaint table on a patio under an umbrella surrounded by a fence. Trees are extending their branches over the seating area to cast a shadow over the seats. | This image depicts a serene backyard patio scene, featuring a round table and chairs beneath a yellow umbrella. The table is adorned with a laptop, book, and glass, suggesting a quiet workspace or reading nook. The patio is surrounded by a wooden fence, with pink flowers blooming on the |

| # | Question category | Image | Question | Annotator response with LLM Assistant | Annotator response without LLM Assistant | LLM Response |
|---|---|---|---|---|---|---|
| | | | | Leafy trees with pink flowers climb higher than the top of the treilised fence.<br><br>Large pink petals color the gray tile that makes up the floor of the patio that extends from side to side. The features of a statue are sharpened by shadow as it stands on a pedestal near the corner of this outdoors space. | This situation looks ideal for anyone looking to eat lunch in the shade on a warm summer day. | bushes and trees, adding a touch of natural beauty to the setting.<br><br>A statue of a woman stands on a pedestal in the background, while a beige shade cloth provides additional shade.<br><br>The image exudes a sense of calmness and tranquility, inviting the viewer to relax and unwind in this peaceful outdoor space. |
| 3 | Object Recognition | 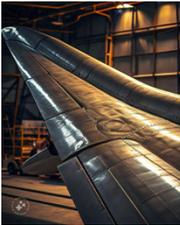 | Describe it. | The picture captures a detailed view of the aircraft wing's underside, where the wing's surface mirrors the environment around it. There is a slight red, white, and grey decal on it. The background seems to be a hanger with wooden pillars supporting the shape of the triangular roof. | Unfortunately I am unable to identify what the object in the picture is. However, it appears to be made out of metal and painted gray with a hatch door attached to one side of the object. | The image shows a close-up of the underside of an aircraft wing, with the wing's surface reflecting the surrounding environment. The wing's surface is a shiny gray metal, with a small section of rivets visible on the right side. The wing's leading edge is curved, and the trailing edge is straight.<br><br>In the background, a warehouse or hangar is visible, with metal beams and a corrugated metal roof. The image appears to have been taken from a low angle, looking up at the wing from the ground. |
| 4 | Object Recognition | 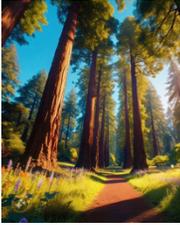 | Tell me what those are. | The trees featured in this image are loblolly pines (Pinus taeda). This species of pine tree is native to the Southeaster United states, in a region stretching from East Texas and Florida north to southern New Jersey. The tree can reach a height of 98-115 feet (30-35 meters) and is known for the scaly plates on its red-brown bark. The loblolly pines are highly valued by the wood industry for their range of uses, including construction materials, lumber and plywood. | This photo showcases a wooded area on a sunny day. The trees themselves are difficult to verify without more context, however, I can describe them as being very tall with green leaves or needles at the very tops of them. | The image appears to show a stand of pine trees. The trees are tall, slender, and have long, needle-like leaves that are a deep green color. They are spaced fairly close together, with some of them growing in clusters or groups. The trees appear to be in a wooded area, possibly a forest or a park, and there is a pathway running through the center of the image. The sky above the trees is blue and cloudy. |
| 5 | Suggestion/Recommendation | 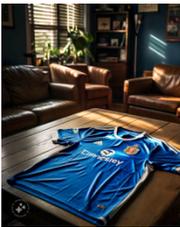 | What does this shirt go with? | The image shows what appears to be a white, soccer jersey with the name "Atajuelense" on it, indicating that the jersey represents a Costa Rican soccer team.<br><br>Whatever suits the user's tastes should be worn with the item but | The black-and-white jersey in this image would go well with black jeans, red or black sweatpants or shorts. | The shirt in the image appears to be a soccer jersey, so it would likely be worn with athletic shorts and shoes. The jersey has the name "Atajuelense" on it, which is a Costa Rican soccer team, so it would likely be worn to play soccer or to watch a soccer game. |

| # | Question category | Image | Question | Annotator response with LLM Assistant | Annotator response without LLM Assistant | LLM Response |
|---|---|---|---|---|---|---|
| | | | | athletic shorts that accompany the jersey along with sneakers may complement the jersey well. | | |

Table A5: Example of LLM Judge disagreement with human researcher feedback

| Query | How often does this need to be watered? |
|---|---|
| Caption | Meet Whorled Milkweed (Asclepias verticillata), a rare gem in New England's fields. Its unique whorled leaves and small flowers make it a standout among milkweeds. #NativePlantTrust #Milkweed #PollinatorPlants |
| Annotator Response | The plant depicted seems to be Asclepias fascicularis, commonly referred to as narrow-leaf milkweed. The narrow-leaf milkweed is a hardy perennial that thrives in various habitats and conditions. This low-maintenance plant is commonly included in wildlife gardens and restoration projects, where it is sown from seeds to blend seamlessly with native grasses and wildflowers. In the spring, following winter rains, they produce green, linear leaves. The ball-shaped clusters of flowers consist of numerous small, white to light pink blossoms that appear as well. To maintain growth, avoid overwatering, overwatering can damage its root system and expose it to fungal infections and other diseases that could harm. |

| | Human Researcher Grade | Human Researcher Explanation | LLM Judge Grade | LLM Judge Explanation |
|---|---|---|---|---|
| Grammar/Presentation | Minimum | The last sentence is a run-on, not grammatically correct; "To maintain growth, avoid overwatering, (could end sentence here, or use semi colon to lead in to next clause) overwatering can damage its root system and expose it to fungal infections and other diseases that could harm." (should finish sentence with subject, "it." | Minimum | The annotator response is well-formatted, with proper sentence structure and no grammatical errors. The response is easy to read and understand, with clear and concise language used throughout. The annotator has also provided relevant information about the plant, including its habitat, growth, and maintenance requirements. The response is well-organized and logically structured, making it easy to follow and comprehend. Overall, the annotator response meets the 'Good' criteria for Grammar & Presentation. |
| Tone/Style | Minimum | The tone/style is okay; it meets minimum requirements like being straightforward but is not comprehensive, and it does not answer the question completely. | Minimum | The annotator response mirrors the fluidity and subtleties of human conversation, sounding natural and human-like. The tone is straightforward and non-judgmental, neither overly friendly nor overly serious. The response is centered on information rather than opinion or emotion, providing comprehensive details about the plant's characteristics and care. The language used is smart and straightforward, directly addressing the query without unnecessary jargon or technical terminology. The response shows empathy for the user's query, providing guidance on how to maintain the plant's growth. Overall, the tone and style of the annotator response meet the 'Good' criteria. |
| Instruction-Following | Insufficient | The response does not fully and explicitly answer the prompt/ask: "How often does this need to be watered?" The response does not say how often, only that the plant should not be overwatered. | Insufficient | The annotator response addresses all explicit asks of the query 'How often does this need to be watered?' by providing information on the watering needs of the narrow-leaf milkweed plant. The response satisfies all parameters or constraints provided within the query, including the context of the image and the country (New England). The annotator has followed the instructions to provide a response that resembles a question a user would ask an AI assistant. The response is clear and concise, providing relevant information about the plant's watering needs and how to maintain its growth. Overall, the annotator response meets the 'Good' criteria for Instruction-Following. |

| Query | How often does this need to be watered? |
|---|---|
| Caption | Meet Whorled Milkweed (Asclepias verticillata), a rare gem in New England's fields. Its unique whorled leaves and small flowers make it a standout among milkweeds. #NativePlantTrust #Milkweed #PollinatorPlants |
| Annotator Response | The plant depicted seems to be Asclepias fascicularis, commonly referred to as narrow-leaf milkweed. The narrow-leaf milkweed is a hardy perennial that thrives in various habitats and conditions. This low-maintenance plant is commonly included in wildlife gardens and restoration projects, where it is sown from seeds to blend seamlessly with native grasses and wildflowers. In the spring, following winter rains, they produce green, linear leaves. The ball-shaped clusters of flowers consist of numerous small, white to light pink blossoms that appear as well. To maintain growth, avoid overwatering, overwatering can damage its root system and expose it to fungal infections and other diseases that could harm. |

| | Human Researcher Grade | Human Researcher Explanation | LLM Judge Grade | LLM Judge Explanation |
|---|---|---|---|---|
| Comprehensive | Insufficient | The response is incorrect; the plant pictured is the whorled milkweed, which was mentioned in the caption as well as easy to find on a side search. Response is significantly lacking in detail and fails to cover essential aspects of the question or request. Because the information is incorrect, it will leave the user with an incomplete or misleading understanding. | Insufficient | I have evaluated the annotator_response based on the provided materials and guidelines. The annotator_response provides a good level of detail, insight, and nuance. It correctly identifies the plant as Asclepias fascicularis, commonly referred to as narrow-leaf milkweed, and provides relevant information about its characteristics, growth habits, and maintenance requirements. The response also includes specific details about the plant's appearance, such as the green, linear leaves and ball-shaped clusters of flowers. Additionally, it provides useful advice on how to maintain the plant, including avoiding overwatering to prevent root damage and diseases. The response is well-organized, coherent, and easy to follow, making it comprehensive and satisfying for the user. Overall, the annotator_response meets the 'Good' criteria for Comprehensive quality. |